\documentclass[preprint,12pt]{iopart}
\usepackage[pdftex]{graphicx}
\usepackage{hyperref}
\pdfoutput=1
\hypersetup{pdfauthor={some author},pdftitle={eye-catching title}}
\ifpdf

\begin{document}

\title{Radiation defects in BaF$_2$-Cd}
\author{A.V. Egranov$^{1,2}$ }

\address{$^1$ Vinogradov Institute of Geochemistry, Russian Academy of Sciences, Favorskii street 1a, 664033 Irkutsk, Russia}
\address{$^2$ Irkutsk State University, Faculty of Physics,  Gagarin Blvd. 20, 664003 Irkutsk, Russia}
\ead{alegra@igc.irk.ru}

\begin{abstract}
Radiation defects in barium fluoride single crystals doped with cadmium have been investigated by luminescence and absorption spectroscopy, as well as by electron  spin resonance spectroscopy. Three types of Cd$^+$ centers differing by the local environment with the point symmetries O$_h$, C$_3$$_v$ and C$_2$$_v$  have been identified although in the crystals only the cubic centers are usually formed. We believe that these features may arise from the difference in the spacial distribution of the impurities in the crystals.
\end{abstract}
% PACS, the Physics and Astronomy % Classification Scheme.

\maketitle

\section{Introduction}
It has been found that alkaline earth fluoride crystals which have not been deliberately doped with impurities are much less susceptible to coloration at room temperature by x - irradiation than most alkali halides. Undoped CaF$_2$ and SrF$_2$ crystals may be colored by x-rays much more readily at 4 K or at 77 K than at room temperature but the coloration efficiency, especially in the case of CaF$_2$ is still much slower than in most alkali halides \cite{Hayes:1974,Call:1974}. The extremely slow coloration rate of undoped alkaline earth fluorides by x-irradiation is due to inefficient separation of close F-H pairs. It has been also found \cite{Hayes:1973,Hayes:1973a} that the holes complementary to the F centers produced by x -irradiation at both 77 and 4 K are V$_k$ centers. In alkali halides, by contrast, the irradiation in the helium temperature range generally produces H centers than V$_k$ centers \cite{Kanzig:1959}.

Introduction of some cationic impurities in alkaline fluoride leads to a significant increase coloration of the crystals by ionized irradiation.  It has been found that in some cases the composition of the radiation defects includes anion vacancies that in general it is unusual, as in the undoped crystals, they are not created with such efficiency. Some propositions on the matter were made by us in the previous papers \cite{Egranov:2008, Egranov:2008a, Egranov:2013}. Influence of cationic impurities on the formation of intrinsic defects in the anion sublattice can be reduced to two essentially different processes:

- The centers, including the impurity ion and anion vacancy are created at the temperatures higher than the onset of the motion of anion vacancies. Onset of the motion of ananion vacancy in CaF$_2$ crystal is at about 200K.  In this case, the formation of anionic vacancies is presumably due to the non-radiative decay of self-trapped excitons \cite{Hayes:1974, Song:1993}. Impurity charge defects stimulate the separation of charged intrinsic ${\alpha}$-I(F$^-$$_i$) defects \cite{Egranov:2008, Egranov:2008a} by their electric fields. The formation of F$_{2A}$$^+$ in CaF$_2$-Na \cite{Tijero:1990}, Cd$^+$(C$_{3v}$) in CaF$_2$ and SrF$_2$ \cite{Egranov:2008, Egranov:2008a} and Mn/F -centers in CaF$_2$-Mn \cite{McKeever:1986} is apparently connected with the motion of anion vacancies.

- The centers, including the impurity ion and anion vacancy are created at 77 K and the formation are not related to the heat-activation process of the motion of anionic vacancies. It seems that formation is linked to the configuration instability at the  impurity ion trapping an electron (maybe hole) \cite{Egranov:2013}. The photochromic centers are produced either by x-irradiation or by additive coloration (by heating the crystals in a calcium atmosphere) of CaF$_2$ crystals doped with certain rare earths ions (La, Ce, Gd, Tb and Lu) or Y \cite{Anderson:1971, Staebler:1971, Alig:1971}. It has been found \cite{Bugaenko:2008,Sizova:2012} that x-irradiation at 77 K of the CaF$_2$ and SrF$_2$ crystals doped with the impurities which can form the photochromic centers, results in creation PC$^+$ and V$_k$ centers (self-trapped hole which have the structure of molecular ion - F$_2$$^-$). The chemical instability of the divalent compounds for these ions \cite{Cotton:2006} leads to the formation of the PC$^+$ centers, instead of the divalent ions in alkaline earth fluorides. 

While the second type of the processes occurs much less frequently than the first, however, it is not limited only the photochromic centers that formed in alkaline earth fluorides activated some rare-earth ions. So in CaF$_2$-Co the defects consisting  of  an  F  center nearest  to  a Co$^{2+}$  ion in  a substitutional position are  created  at  300 and 80 K \cite{Alcala:1978}.

In crystals of calcium  and strontium fluorides activated divalent ions of cadmium, radiation leads to formation of monovalent cadmium centers perturbed by one or two anion vacancies, located in the immediate environment, which lowers the center of symmetry to C$_{3v}$ and C$_{2v}$, respectively  \cite{Egranov:2008, Egranov:2008a}. Usually in crystals of barium fluoride  Cd$^+$ centers in a cubic environment are only created. However, in some cases it is possible to produce in these crystals the centers similar to those obtained in other alkaline earth fluorides such as CaF$_2$ and SrF$_2$ \cite{Springis:1996}. In this article radiation defects in barium fluoride single crystals doped with cadmium have been investigated by luminescence and absorption spectroscopy, as well as by electron  spin resonance spectroscopy. Three types of Cd$^+$ centers differing by the local environment with the point symmetries O$_h$, C$_3$$_v$ and C$_2$$_v$ have been identified. We discuss the reasons for the formation of Cd$^+$ centers with low symmetry in these crystals.  

\section{Experimental technique}
The crystals of BaF$_2$ doped with cadmium (with the concentration up to 1mol \%) were grown from the melt by the Bridgman-Stockbarger method. In alkaline earth fluorides single crystal growth a small amount of CdF$_2$ are generally used as a scavenger in order to remove oxides and oxyfluorides contained in the raw materials by the reaction  CdF$_2$ + BaO${\rightarrow}$ CdO + BaF$_2$. CdO and excess CdF$_2$ evaporate completely from the melt before crystallization begins \cite{Stockbarger:1949,Sobolev:2002}. Because of the appreciable vapor pressure of CdF$_2$  at the melting point of BaF$_2$  it was necessary to confine the impurity-doped melt in the crucible by means of a floating graphite plug, which reduced the open surface area of the melt and thus decreased the impurity evaporation. The samples were of high optical quality and no indication of oxygen contamination. The atomic emission spectrometry analyses were carried out on severals samples in order to determine the concentrations of Cd  in the grown crystals. Optical absorption measurements were made with a Lambda 950 UV/VIS/NIR  spectrophotometer. Luminescence measurements were made with a LS 55 Luminescence Spectrometer Perkin Elmer with PMP R928. The crystals were irradiated at 77 or 295 K by x-rays from a Pd tube operating at 50 kV and 50 mA during not more than one hour. The EPR experiments were done on an X-band spectrometer Fourier Transform EPR Spectrometer ELEXSYS Series Bruker. 

\begin{figure}
\centering
\includegraphics[width=5.2in, height=3.6in]{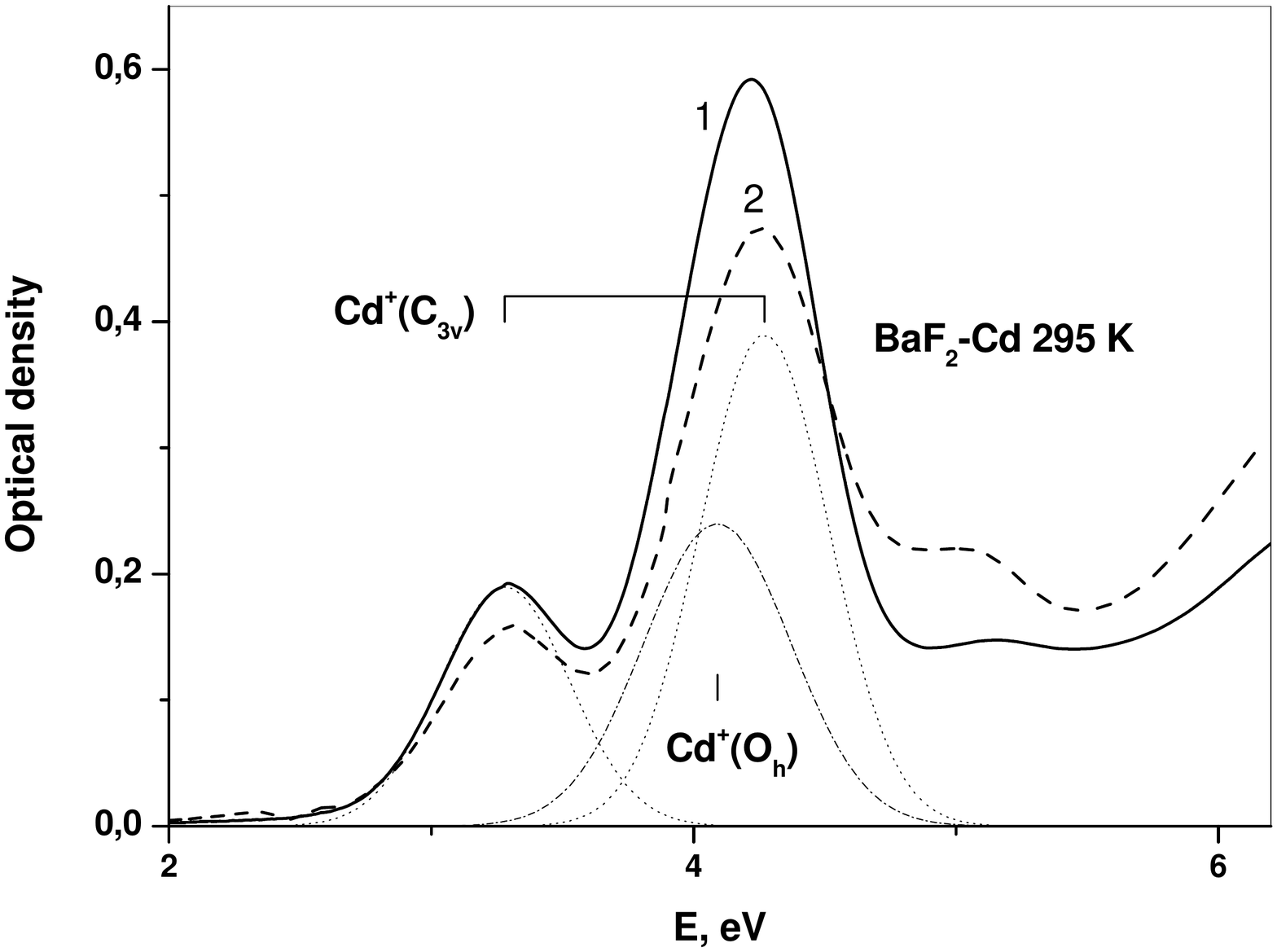}
\caption{Spectra of optical absorption at 295K (1,2) of BaF$_2$-Cd crystals after x-irradiation at 295K (1) and subsequent
optical bleaching by 3.7 eV light at 295K (2). Dot curves are Gaussian bands to simulate measured spectrum.}
\label{absorption}
\end{figure}

\section{Experimental result}
\subsection{Absorption}
X-irradiation at 77 K of BaF$_2$ crystals doped with cadmium results in the creation of absorption bands at 4.1 eV of reduced cadmium centers with cubic symmetry - Cd$^+$(O$_h$)  \cite{Nepomnyashchikh:2005} and 3.4 eV of V$_k$ centers \cite{Nepomnyashchikh:2002}. In some crystals the Cd$^+$ centers with cubic symmetry created by x-irradiation at 77 K are converted to Cd$^+$ centers with lower symmetry by heating up to room temperature. X-irradiation at room temperature of BaF$_2$-Cd crystals leads to the same results. There are two absorption bands with peaks at 3.28 and 4.27 eV with an approximate intensity ratio of 1:2 in the  absorption spectra of the crystals BaF$_2$-Cd after  x-ray irradiation at room temperature (Fig.~\ref{absorption}). By analogy with the results obtained for for other alkaline earth fluorides \cite{Egranov:2008, Egranov:2008a}, these bands are assigned to Cd$^+$ centers which consists of a fluorine vacancy and the nearest-neighboring Cd$^+$ ion. In this case the symmetry center is reduced to C$_{3v}$  (Cd$^+$(C$_{3v}$)), which determines the presence of two absorption bands. The centers are optically bleached by light in the second absorption band with a maximum at 4.27 eV and the high-energy absorption at about 5.0 eV is increased, which associated with the formation of cadmium centers Cd$^+$(C$_{2v}$) (by analogy to the previous results \cite{Egranov:2008, Egranov:2008a}) (Fig.~\ref{absorption}). 

\begin{figure}
\centering
\includegraphics[width=5.2in, height=3.6in]{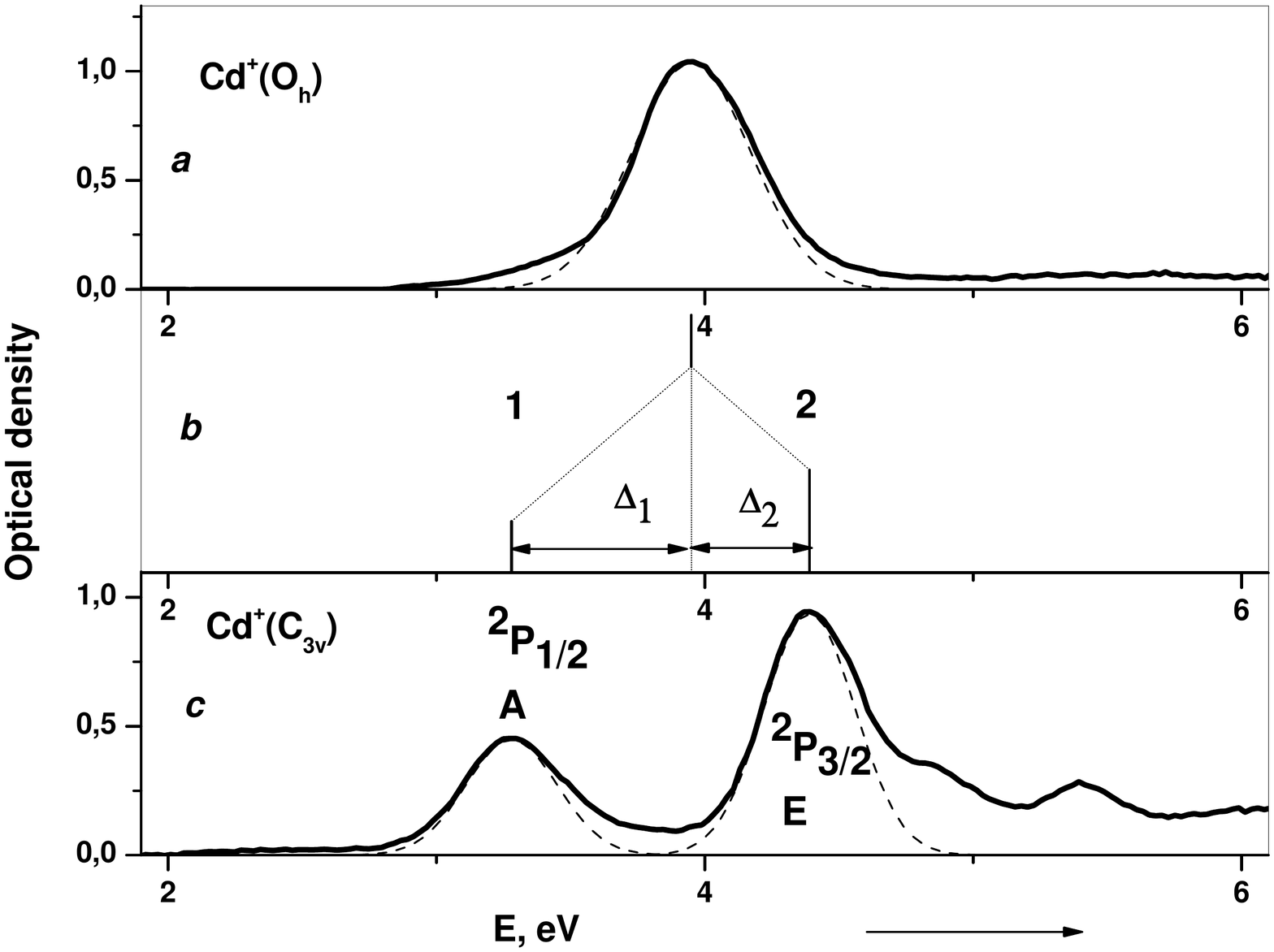}
\caption{Schematic representation of the level splitting Cd$^+$ ions under lower symmetry.
a) The absorption spectrum of Cd$^+$(O$_h$) ions in cubic environment, b) {$\Delta$}$_1$ and {$\Delta$}$_2$ shift of the absorption bands of Cd$^+$(C$_{3v}$) centers relative to the absorption band of Cd$^+$(O$_h$) centers c) the absorption spectrum of Cd$^+$(C$_{3v}$) ions, which are localized near the anion vacancy}
\label{schema}
\end{figure}

Experimental data on the splitting of the absorption band of Cd$^+$(C$_{3v}$) centers in alkaline-earth fluorides (Fig.~\ref{schema})) are presented in the table 1. From the table it is clear that the splitting of the p-state  of Cd$^+$(C$_{3v}$) center in all crystals has approximately the same value of about 1 eV. In this case, ${\Delta}$$_1$ is monotonically increases from CaF$_2$ to BaF$_2$ and ${\Delta}$$_2$ has the opposite behavior.

\begin{table}
\caption{\label{tab:table1}Splitting of absorption band of the Cd$^+$(C$_{3v}$) centers in alkaline earth fluorides (for difinitions see Fig.~\ref{schema})}
\begin{tabular}{ccccc}
Crystal & ${\Delta}$$_1$  & ${\Delta}$$_2$ & ${\Delta}$$_0$=${\Delta}$$_1$+${\Delta}$$_2$ &  T, K\\
 \hline
CaF$_2$ & 0.17 & 0,73& 0.9 & 295\\
  & 0.25&0.73 & 0.98 &80\\
SrF$_2$ & 0.65&0,44 &1.09 & 295\\
 &0.67&0.44& 1.11&80\\
 BaF$_2$&0.8&0.19&0.99& 295\\
 &0.85&0.15&1.0&80\\
\end{tabular}
\end{table}

\subsection{EPR}
\begin{figure}
\centering
\includegraphics[width=5.2in, height=3.6in]{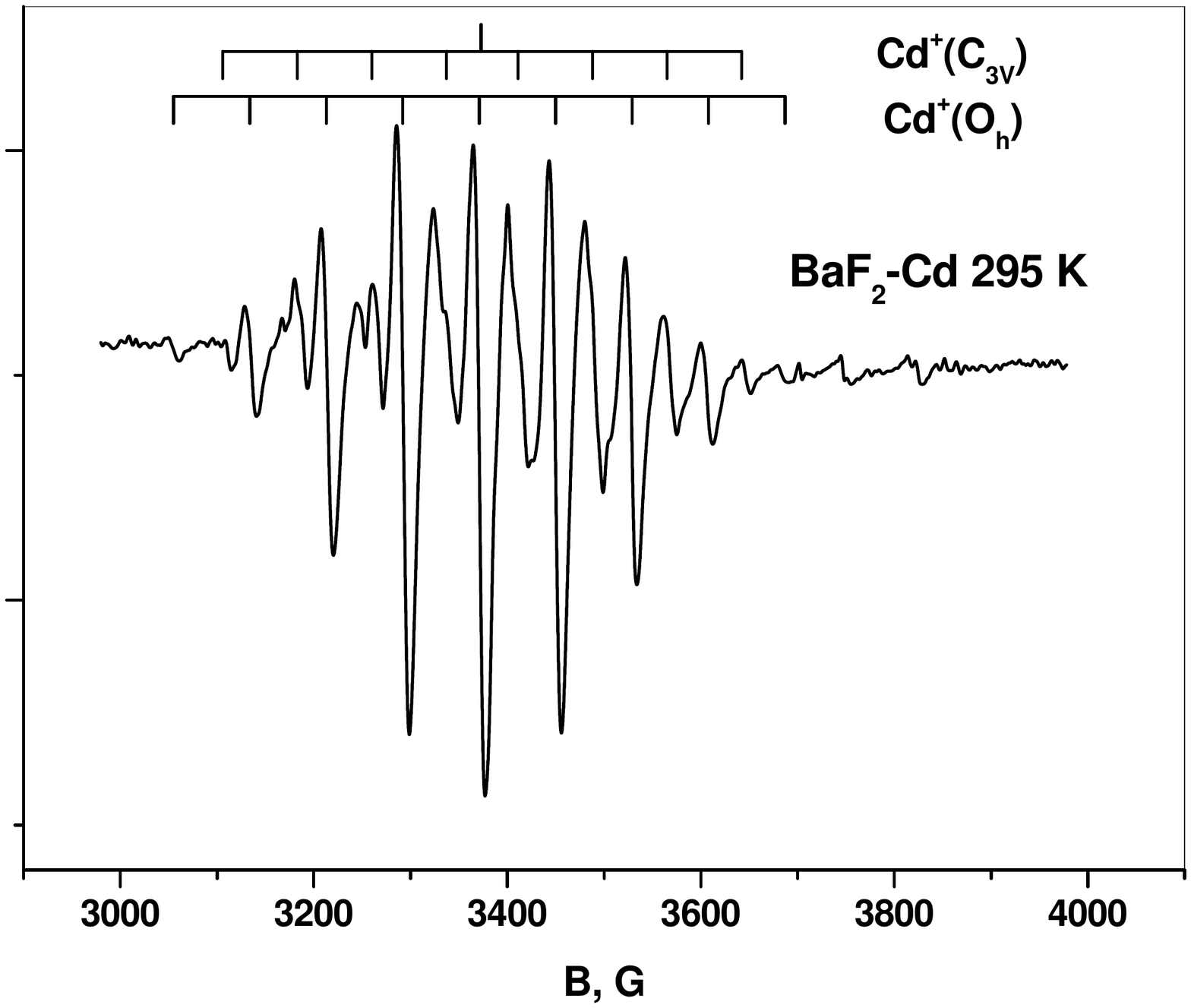}
\caption{EPR spectrum of  BaF$_2$-Cd crystal after x-irradiation at 295 K, B{$\parallel$}$<$100$>$.}
\label{epr}
\end{figure}

The EPR spectrum of BaF$_2$-Cd crystal x-irradiated at 295 K is shown in Fig.~\ref{epr}.   There are two groups of lines which are related to the centers of monovalent cadmium ion with different symmetry. More intense group consisting of nine lines refers to the ions Cd$^+$(O$_h$) in a cubic environment. The spectrum is due to superhyperfine interaction of the unpaired electron with eight equivalent fluorine nuclei. This is similar to what was observed earlier \cite{Nepomnyashchikh:2005, Krutikov:1976}. Another group with eight lines is the EPR spectrum of monovalent ions of cadmium Cd$^+$(C$_{3v}$), with adjacent anion vacancy, as indicated by the absence of one fluorine nuclei in this spectrum. In addition there is poor resolution structure in each line, which is not allowed at the temperature of liquid nitrogen. Although the centers are oriented along the C$_3$ axis, the simplest structure is observed for B{$\parallel$}$<$100$>$, since the angles between eight different positions of the anion vacancy on the tops of the cube, the electron and the direction of the external magnetic field are the same. The fact that the cadmium ion does not experience a significant shift from the center indicates similar values  of superhyperfine splitting for the ions in the cubic environment and disturbed anion vacancy. It should be noted that although the EPR signal from Cd$^+$(O$_h$) more intense than that of Cd$^+$(C$_{3v}$) centers absorption spectra show that the concentration of Cd$^+$(C$_{3v}$) centers higher than that of the Cd$^+$(O$_h$) centers.

\subsection{Luminescence}
\begin{figure}
\centering
\includegraphics[width=5.2in, height=3.6in]{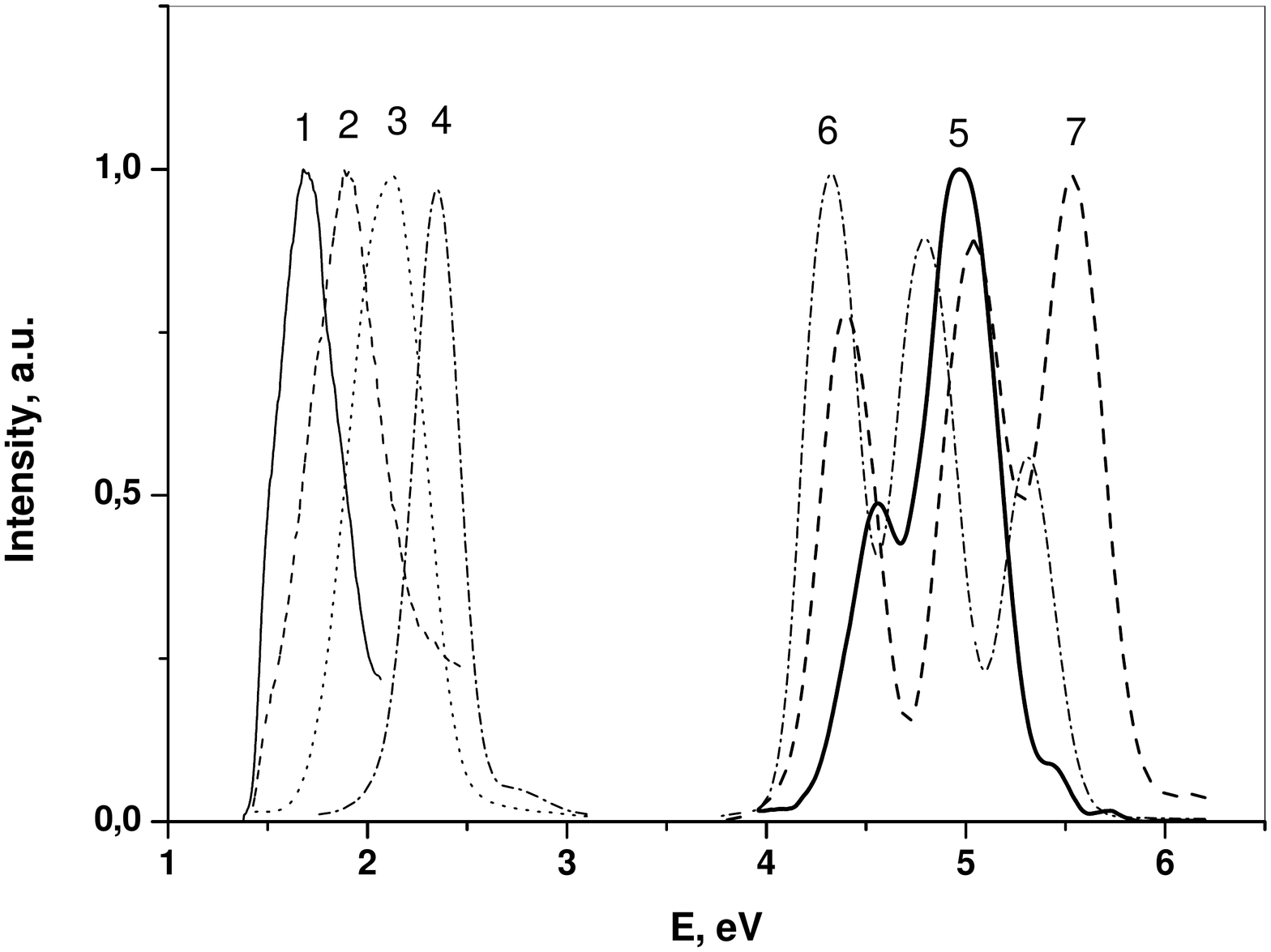}
\caption{Spectra of emission (1,2)  and excitation (5) of BaF$_2$-Cd crystals  at 80 (1,5) and 295 K (2) after X-ray irradiation at 295 K and subsequent  bleaching with light 3.7 eV at 295 K. For comparison the spectra of excitation (6,7) and emission (3,4) of Cd$^+$(C$_{2v}$)  centers in crystals of CaF$_2$ (4,7) and SrF$_2$ (3,6) at 77 K show}
\label{emission}
\end{figure}

The figure~\ref{emission} shows the emission and excitation spectra of Cd$^+$(C$_{2v}$) centers in BaF$_2$ crystals  compared with the spectra of similar centers in other alkaline earth fluorides. Photoexcitation at room temperature in the absorption band of Cd$^+$(C$_{2v}$) centers at about 5.0 eV (Fig.\ref{absorption}), which is formed in BaF$_2$-Cd crystals by x-irradiation and subsequent optical bleaching with 3.7 eV light at 295K,  leads to the appearance of red emission with peak at 1.9 eV, which shifts to 1,7 eV  at 77 K. Unlike the Cd$^+$(C$_{3v}$) centers for which the splitting of p-state changes little on going from CaF$_2$ to BaF$_2$ for Cd$^+$(C$_{2v}$) centers largest splitting is observed in crystals CaF$_2$, and the smallest in BaF$_2$ crystals. This leads to the fact that, unlike CaF$_2$ and SrF$_2$, the two excitation bands  are only  observed in BaF$_2$ crystals, the intensity of the high energy band twice higher than that of the low energy band, indicating that the former arises from the two unresolved transitions to the excited p state.

\subsection{Temperature dependence}
Figure 5 shows the temperature dependence of the formation of the centers having in its composition anionic vacancy. In the case of formation of F$_2$$^+$(Na) centers in CaF$_2$ the dependence is close to that reported by Tijero et al \cite{Tijero:1990}. In agreement with Tijero et al \cite{Tijero:1990}, the experimental data obtained from  x-ray-irradiated at 77 K CaF$_2$-Na and SrF$_2$-Na crystals show that the (F$_2$$^+$)$_A$ centers are created in two stages: the low-temperature stage is associated with a short-range reorientation, while an anion-vacancy-diffusion process operates in the high-temperature zone. There is some distinction between the temperature range of the formation of F$_{2A}$$^+$  and Cd$^+$(C${_3v}$) centers. Apparently temperature dependence is not caused  by the motion of free anion vacancies but probably results from motion within the Coulomb interaction impurity-vacancy pairs.

\begin{figure}
\centering
\includegraphics[width=5.2in, height=3.6in]{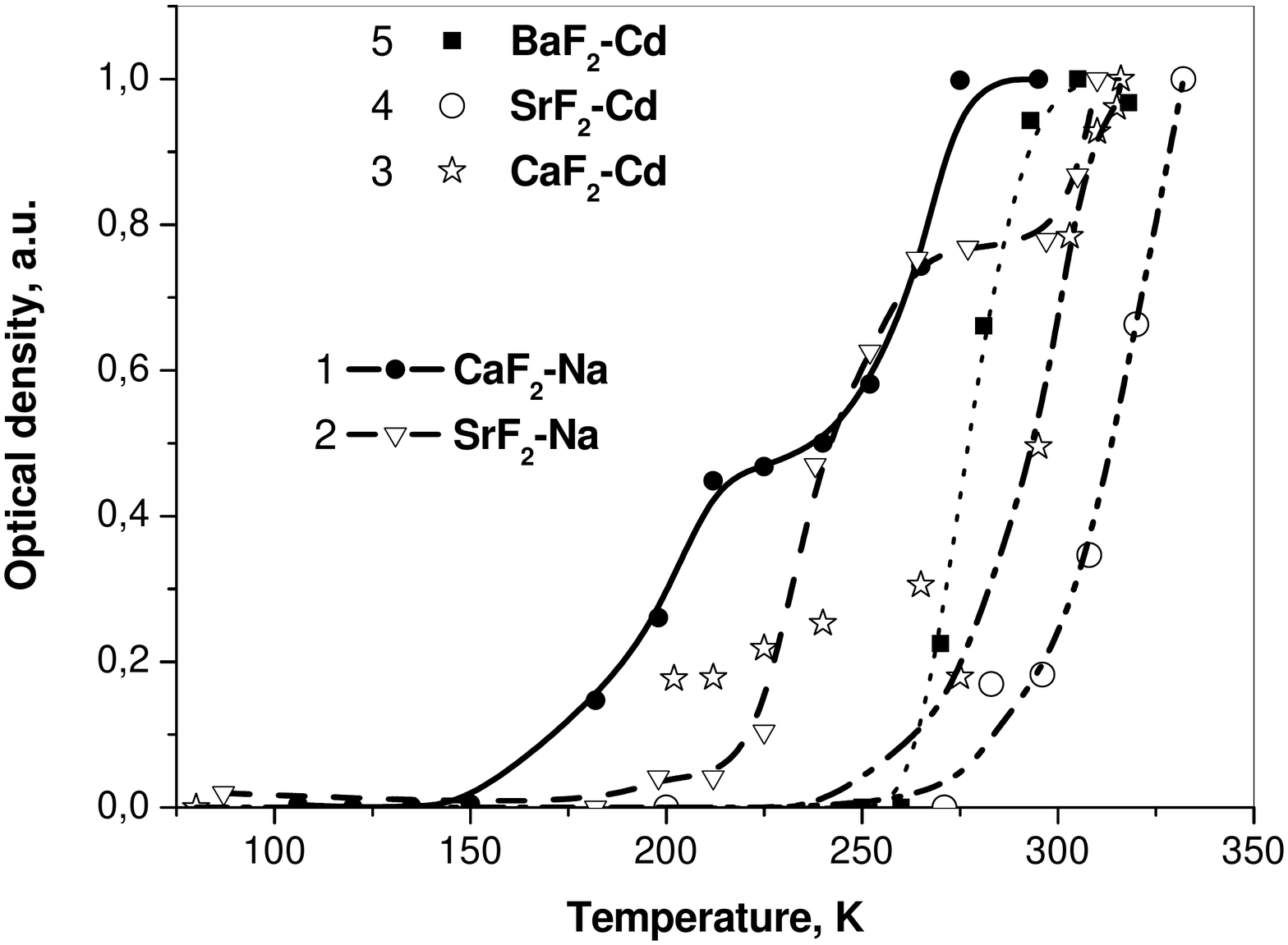}
\caption{Temperature dependence of the formation of F$_2$$^+$(Na) centers in CaF$_2$-Na (1), SrF$_2$-Na (2) crystals and Cd$^+$(C$_{3v}$) centers in CaF$_2$-Cd (3), SrF$_2$-Cd (4), BaF$_2$-Cd (5) crystals. Measurements were carried out at 77 K. The samples were held for 2 min at each temperature.} 
\label{temperature}
\end{figure}

\section{Discussion}
In our previous papers \cite{Egranov:2008,Egranov:2008a}  three types of Cd$^+$ or Zn$^+$ centers differing by the local environment with the point symmetries O$_h$, C$_3$$_v$ and C$_2$$_v$ have been identified in CaF$_2$ and SrF$_2$ crystals. The formation of the last two results of the interaction between the reduced impurity centers and anion vacancies. The latter are intrinsic radiation defects and are not essentially created by x - irradiation in the undoped crystals. It has been suggested that under x-ray radiation  of the crystals charged defects such as Cd$^+$ or Zn$^+$ in cubic symmetry sites are created and their electric field  promotes the separation of intrinsic radiation defects (${\alpha}$-I pairs) formed by radiationless decay of self-trapped excitons near the charged impurity centers. Radiation formation of the anion vacancies occurs, apparently, near the charged impurity defect, at distance of several constant lattices (because the next process of the formation of the imputiry-anion vacancy complex is thermally activated). Formation of the anion vacancies occurs owing to two competing processes. On the one hand this is the random formation of an exciton near the charged impurity defect which probability increases with increasing the number of anion sites, so with increasing the distance from the impurity defect. On the other hand the strength of electric field which promotes the separation of ${\alpha}$-I pairs decreases with increasing the distance from the charged impurity ion. It is necessary to make the remark, although there is no doubt that the F$_{2A}$$^+$(Na) and Cd$^+$(C$_{3v}$) centers include anion vacancy, there are no F centers (except F$_A$ in the Na doped cryctals) and attempt to produce from these anion vacancies F - centers by additional radiation in the temperature range from 77 to 300 K the crystals previously irradiated at 77 K fails. This indicates that these anionic vacancies can not be considered as free but as interacting with the charged impurity ion, even at 77 K.

At the same time the strength of electric field is determined by distribution of the charged impurity defects in crystals. Low concentrations of charged defects  in crystals generate random electric fields characterized by Holtsmark distributions \cite{Holtsmark:1919,Redfield:1963,Redfield:1964}. This case was described in our previous articles \cite{Egranov:2008,Egranov:2008a} and efficiency of the formation of the Cd$^+$(C$_{3v}$) centers having in its composition anionic vacancy decreases from CaF$_2$ to SrF$_2$ and in BaF$_2$ crystals the centers are not created under x-irradiation at room temperature.  In most cases in barium fluoride crystals doped with divalent ions of cadmium, x-ray irradiation at room temperature leads to the formation of Cd$^+$(O$_h$) centers in cubic environment.

With the increase of concentration of  impurity distribution is constantly changing from random to uniform as the number of crystal lattice sites are limited. At this uniform distribution, by symmetry, the electric field is zero everywhere and separation ${\alpha}$-I pairs does not occur and anion vacancies are not created. As a result Zn$^+$(O$_h$)  centers in a cubic environment are  only created after x-irradiation of SrF$_2$ doped with at about 1 mol \% ZnF$_2$ at room temperature. At lower concentration the impurity centers incorporating anion vacancy such as Zn$^+$(C$_{3v}$) and Zn$^+$(C$_{2v}$) are observed.

By the difference in the conditions of crystal growth non-uniform distribution of impurity in crystal is also possible. In this case it is expected the region with higher strength of electric field in comparison with random distribution. Two types of the  BaF$_2$-Cd crystals have been produced - "small" with a diameter of 10 mm and 50 mm length and "large" with a diameter of 60 mm and 150 mm length. In the latter strongly non-uniform distribution of cadmium in the crystal  is observed from time to time. Cd$^+$ centers with lower symmetry in BaF$_2$-Cd are produced in "large" crystals, in which there is a strong non-uniform distribution of cadmium.

\section{Conclusion}
Experimental results received in this and the previous works \cite{Egranov:2008,Egranov:2008a} can be interpreted as follows. The impurity distribution in the crystals is essential for the separation of the intrinsic radiation defects (${\alpha}$-I pairs). Three main types of spacial distribution of impurity in crystal can be identified in oder of increasing strength of internal electric fields of the charged defects - uniform, random and non-uniform. Usually at low concentration of impurity random distribution is realized. In this case efficiency of the formation of the Cd$^+$(C$_{3v}$) and Cd$^+$(C$_{2v}$) centers having in its composition one or two anion vacancies decreases from CaF$_2$ to SrF$_2$ and in BaF$_2$ crystals the centers are not created under x-irradiation at room temperature.  In most cases in barium fluoride crystals doped with divalent ions of cadmium, x-ray irradiation at room temperature leads to the formation of Cd$^+$(O$_h$) centers in cubic environment. However at strong non-uniform distribution of the impurity in BaF$_2$-Cd crystals the centers with low symmetry are also created. The spacial distribution aspires to uniform with incresing concentration of impurity. In this case  the centers  with a cubic symmetry are only created and this process begins in crystals SrF$_2$ standing between CaF$_2$ and BaF$_2$.
 
\section*{Acknowledgments}
This work was partially supported  by The Ministry of education and science of Russian Federation.

\section*{References}
\bibliographystyle{utphys}
\bibliography{Egranov}

\end{document}